# Electronic functionalization of the surface of organic semiconductors with self-assembled monolayers

Feb. 1st, 2007

M. F. Calhoun, J. Sanchez, D. Olaya, M. E. Gershenson and V. Podzorov*

*Department of Physics and Astronomy, Rutgers University, Piscataway, New Jersey, USA*

*Electronic mail: podzorov@physics.rutgers.edu

**Molecular self-assembly has been extensively used for surface modification of metals and oxides for a variety of applications, including molecular[1,2,3,4] and organic electronics[5,6,7,8]. One of the goals of this research is to learn how the electronic properties of these surfaces can be modified by self-assembled monolayers (SAM). Here, we demonstrate a new type of molecular self-assembly: the growth of organosilane SAMs at the surface of organic semiconductors, which results in a dramatic increase of the surface conductivity of organic materials. For organosilane molecules with a large dipole moment, SAM-induced surface conductivity of organic molecular crystals approaches $10^{-5}$ S per square, which is comparable to the highest conductivity realized in organic field-effect transistors (OFETs) at ultra-high densities of charge carriers[9,10,11]. SAM-functionalized organic surfaces are fully accessible to the environment which makes them very attractive for sensing applications. We have observed that the interaction of vapors of polar molecules with SAM-functionalized organic semiconductors results in a fast and reversible change of the conductivity, proportional to the pressure of an analyte vapor.**

In this Letter we report on experiments in which the surfaces of small-molecule organic semiconductors were used as substrates for the self-assembly of different organosilane molecules. Most of the results reported below have been obtained with as-grown *p*-type organic single crystals (rubrene, tetracene) with electrical contacts deposited on their surface by thermal evaporation of silver or application of a thin layer of colloidal graphite. The details of sample preparation can be found elsewhere (see, e.g., [12],[13]). The investigated SAM molecules include several members of fluorinated and non-fluorinated trichlorosilanes, such as, e.g., (tridecafluoro-1,1,2,2-tetrahydrooctyl)trichlorosilane (FTS) and n-octadecyltrichlorosilane (OTS)[14]. Non-silane molecules with a structure similar to that of trichlorosilanes have also been tested in order to identify the functional groups responsible for binding the SAM molecules to organic surfaces (see the Supporting Online Materials (SOM)). Our preliminary analysis indicates that such groups are $SiCl_3$, $Si(OH)_3$ or $COH$, rather than alkyl or fluoroalkyl tails of organosilane molecules. This conclusion is also confirmed by our preliminary data on infra-red spectroscopy of SAM-functionalized organic crystals.

SAM growth has been performed in a setup shown in Fig. 1. A *p*-type organic crystal with pre-fabricated contacts is placed in a vacuum chamber separated by a valve from a small container (volume $\leq$ 1 cm$^3$) with a few drops of concentrated SAM molecules. The chamber is evacuated by a mechanical pump, while the SAM valve is closed. When the pressure reaches $10^{-3}$ - $10^{-2}$ Torr, the pump is disconnected from the chamber, and the SAM valve is slowly opened. Figure 1 shows the dynamics of the current, $I(t)$, flowing through the sample at a fixed applied voltage, $V$. As a result of SAM growth at the sample surface, the current rapidly increases by several orders of magnitude. The SAM-induced surface conductivity, $\sigma$, is non-volatile: the high-$\sigma$ state of the samples with graphite contacts persists after storing the samples under high vacuum, in pure oxygen, or under illumination in humid air. The samples with thin-film silver contacts are stable in vacuum or dry gases, but exhibit a drop in $\sigma$ proportional to the humidity when exposed to humid air (see SOM). Remarkably, we have not been able to observe the ubiquitous "gauge effect"[15] with organic samples fully coated with SAMs.

Although the detailed characterization of SAM-induced conductivity is beyond the scope of this Letter, it is worth mentioning that $\sigma$ exhibits an anisotropy consistent with the molecular packing in rubrene crystals, it is weakly temperature dependent in the range 5 – 300 K, and a metallic $\sigma(T)$ dependence (i.e., $d\sigma/dT < 0$) is observed above 250 K[16] indicating that this system is very interesting for charge carrier transport studies. These characteristics strongly suggest that SAM-induced conductivity is not due to chemical doping of the organic semiconductor bulk, but rather due to a partial transfer of electron density from molecules at the surface of organic semiconductor to the electronegative SAM molecules that occurs as a result of SAM bonding to the surface. The degree of the charge transfer and therefore the conductivity of SAM-functionalized samples is likely to be related to the dipole moment of organosilane molecules. A combination of the surface binding and the charge transfer creates a stable aligned layer of negatively charged SAM molecules immobilized above the SAM/crystal interface and a corresponding layer of mobile holes below the interface (see the model depicted in Fig. 1). The observation of a fast and reversible sensing effect (see below) also speaks in a favor of this model, as opposed to chemical doping.

In order to obtain further insights into the process of self-assembly of organosilanes at the surface of organic crystals we have performed scanning electron microscopy (SEM) and atomic force microscopy (AFM) studies of SAM-functionalized rubrene. Figure 2 shows SEM images of a sample at different stages of SAM growth. These stages approximately correspond to the sample conductivity marked by the letters *a*, *b*, *c* and *d* in Fig. 1. At the initial growth stage, SAM regions (the islands of a lighter shade in Fig. 2(a)) nucleate around surface defects, such as dust or graphite particles: SEM images taken at the initial growth stages reveal a higher nucleation density in the vicinity of the graphite contacts than in the middle of the channel. The islands grow in size as the treatment time increases and eventually coalesce into larger connected patches (Fig. 2(b-c)). At long treatment times (> 10 h), the entire surface of the crystal appears uniformly coated in SEM; this state corresponds to the highest conductivity of the sample. A uniform grayscale shade for different islands and within each island suggests that all the SAM regions are similar in surface properties and thickness. The observed SEM contrast is remarkably high for a pristine organic sample with islands of molecular-scale thickness at the surface; this suggests that SEM might be a powerful tool for studying SAM-functionalized organic semiconductors.

Since electron microscopy cannot be used to measure the SAM thickness, we have performed AFM studies of a partially coated rubrene in order to confirm that the SAM thickness is consistent with the length of trichlorosilane molecules (Fig. 3). The AFM images of the SAM at the surface of rubrene have the following characteristics: (1) the SAM islands can easily be distinguished from other structural features of the rubrene surface, such as molecular growth steps, due to the very distinct morphology (see Fig. 3(a): "thin wavy lines" – the growth steps, "smeared patches" – SAM-coated regions); (2) the surface of SAM-coated regions typically appears more fuzzy in AFM than a clean surface of rubrene; and (3) the average height of the SAM islands is <*h*> = 1.3 ± 0.2 nm, which is in a good agreement with the length of FTS molecule[17].

Interestingly, we have observed that SAM-induced conductivity is very sensitive to the presence of molecular species in the environment. The interaction of SAM-functionalized organic surface with vapors of molecules such as acetone, propanol, ethanol, xylene, chlorobenzene, and water results in a fast and reversible change of σ (Figs. 4 and SOM1). This is in sharp contrast to the behavior of untreated single crystals which are not sensitive to these vapors, as found in our experiments with the "air-gap" OFETs[15]. Figure 4 shows the relative change of conductivity, $\Delta\sigma/\sigma_0$, of rubrene samples functionalized with two different SAMs (FTS and OTS) as a function of the relative vapor density of acetone and propanol in the test chamber, $p/p_0$ ($p_0$ is the saturated vapor density). Figure SOM1 shows the device response to other analytes. Upon exposure to an analyte vapor, the conductivity of the sample rapidly decreases or increases, depending on the type of analyte. Most analytes used in this study (with the exception of xylene) caused a decrease of σ, and analyte molecules with a larger dipole moment produce a stronger effect. The response time is typically a fraction of a second. For instance, exposure of FTS-functionalized rubrene to a saturated vapor of acetone results in an instantaneous 100%-change of the electrical resistivity of the sample (the upper inset in Fig. 4: note the semi-log scale and the sharpness of the changes in $I(t)$). Small increments of an analyte vapor density, $\Delta p < p_0$, result in step-wise changes of σ (the lower inset in Fig. 4). Overall, the conductivity varies linearly with the analyte pressure: $\sigma = \sigma_0 + \alpha \cdot p$, where the coefficient of proportionality α depends on the type of SAM and the analyte. The effect is reversible: initial conductivity $\sigma_0$ can be fully restored by purging the test chamber with a pure nitrogen gas or by pumping. We have found that for a stable and reversible operation of these devices a complete SAM coverage of the crystal surface is necessary.

Although much more work is required to elucidate the microscopic mechanism of the sensing effect in SAM-functionalized organic samples, we speculate that the interaction of polar molecules with the polarized SAM layer results in a modification of the average SAM's dipole moment and a subsequent change of σ. This mechanism differs from the operating principle of organic thin-film transistor sensors, based on the analyte diffusion through the grain boundaries and an increase of the trap density at the organic-dielectric interface[18].

SAM-functionalized organic semiconductors provide an intriguing opportunity for developing sensors with chemical selectivity. Indeed, (**a**) the flexibility of SAM synthesis could allow for tailoring the head groups of SAM molecules to selectively bind to certain analytes, without changing the substrate-binding group; (**b**) the dynamics and even the sign of the sensing effect depend on the analyte (Fig. SOM1); (**c**) the linearity of the device response and the slope which is specific to SAM and analyte might have a potential for selective recognition of a range of analytes with a *composite sensor*, based on two or more devices functionalized with different SAMs (e.g., FTS, OTS, etc.), and, finally, (**d**) the response of these devices to certain analytes depends on the type of electrical contacts: e.g., devices with graphite contacts are not sensitive to $H_2O$ vapor, while those with silver contacts are (Fig. SOM1). Further work is necessary in this direction.

To conclude, we have observed molecular self-assembly of organosilanes at the surface of organic semiconductors that induces a highly conducting surface state of these materials. The interaction of SAM-functionalized organic surfaces with molecular species in the environment results in a rapid and reversible variation of the electrical conductivity of these samples. All these findings indicate that SAM-functionalized organic semiconductors represent a perspective platform for research in molecular self-assembly, charge carrier transport and chemical sensing.

This work has been supported by the NSF grants DMR-0405208 and ECS-0437932.

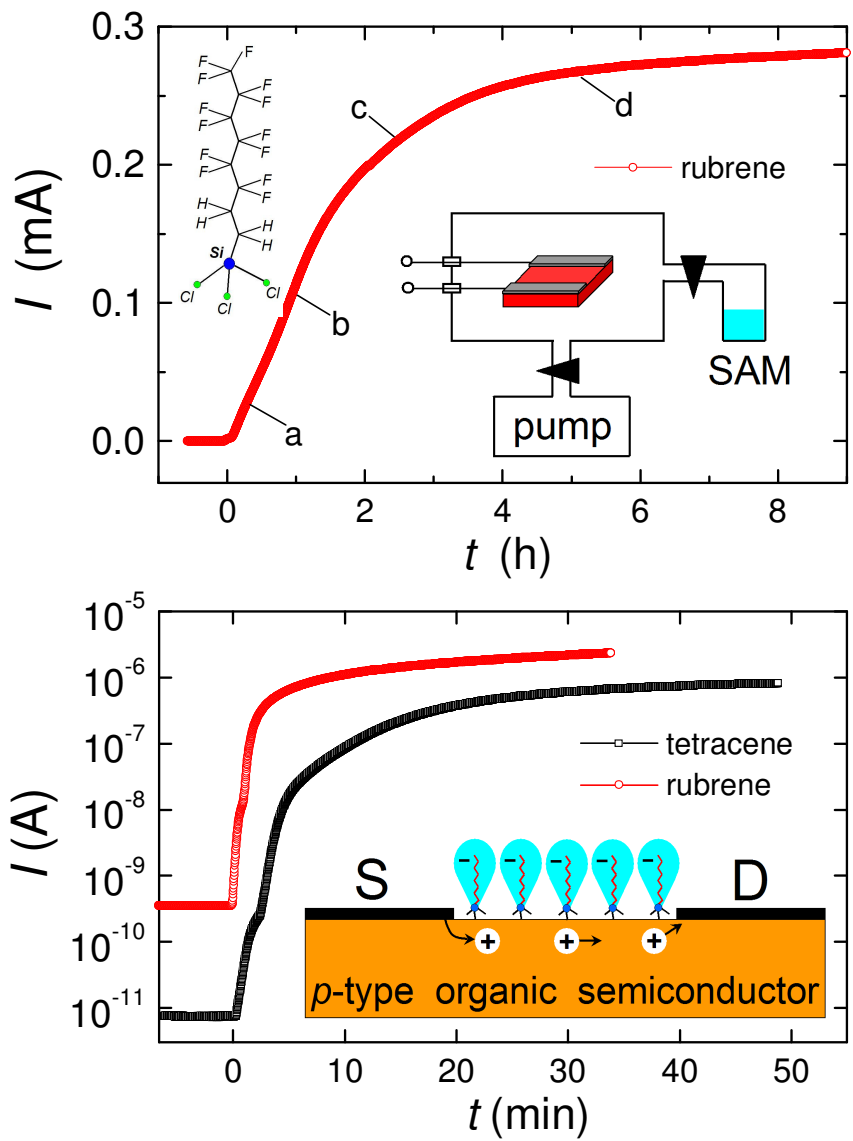

**Fig. 1.** The current flowing at the surface of organic molecular crystals as a function of time, $I(t)$, with $V = 5$ V applied between the contacts. At $t = 0$ the samples are exposed to fumes of fluorinated organosilane: (tridecafluoro-1,1,2,2-tetrahydrooctyl)trichlorosilane. As a result of SAM formation, the surface conductivity of organic crystals increases by 4-5 orders of magnitude. **Top:** $I(t)$ for a rubrene crystal with thermally evaporated silver contacts (channel width and length are $W = 1$ mm, $L = 50$ μm). The letters *a, b, c,* and *d* mark the stages of the treatment, at which the SEM images in Fig. 2 have been taken. The insets on the left and on the right show the structure of the SAM molecule and a diagram of the experimental setup, respectively. **Bottom:** Semi-log $I(t)$ plots for a rubrene (red) and a tetracene (black) samples with colloidal graphite contacts (typical dimensions of the channel are $W = 0.3 - 0.5$ mm, $L = 2.5 - 3$ mm). The inset is a schematic representation of a layer of mobile holes induced by the SAM.

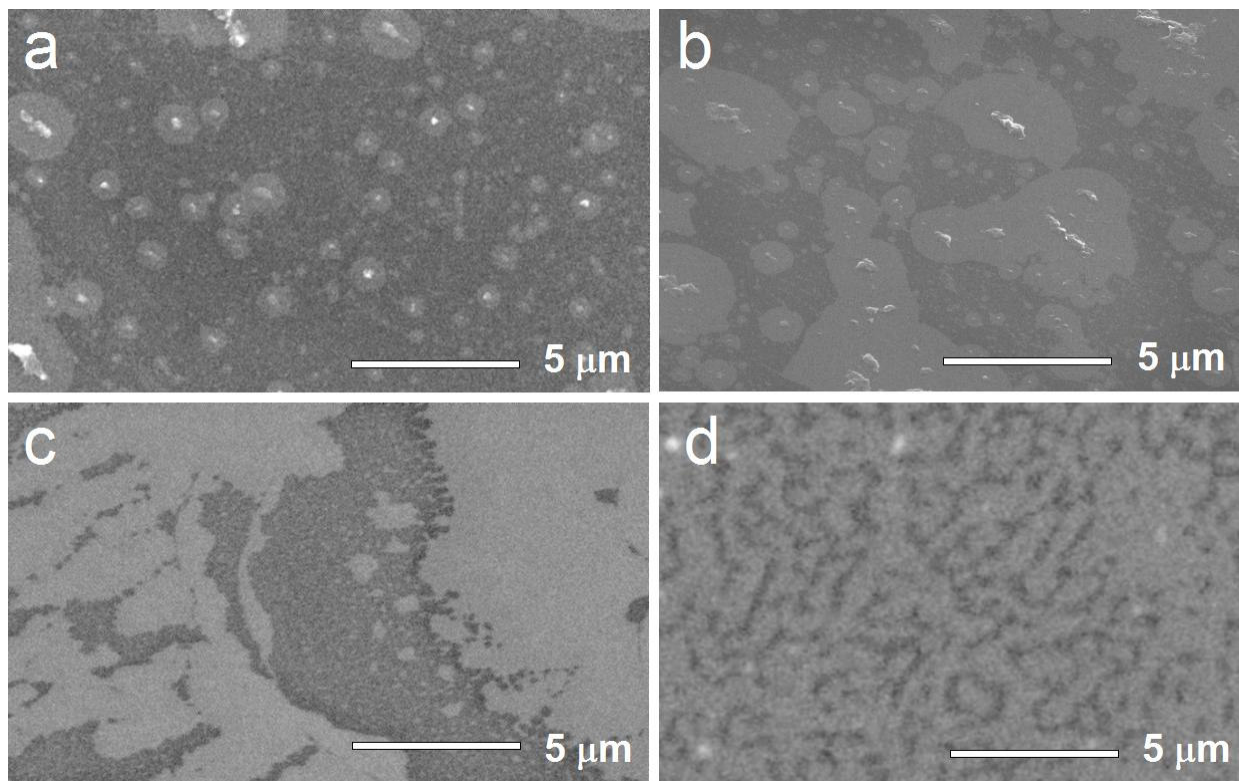

**Fig. 2.** Scanning electron microscopy images of the surface of an (***a,b***)-facet of a rubrene single crystal treated in (tridecafluoro-1,1,2,2-tetrahydrooctyl)trichlorosilane vapors for different periods of time: (a) 15 min, (b) 1 hour, (c) 2.5 hours, and (d) ~ 5.5 hours. The SEM contrast reveals two types of surface: coated with the SAM and uncoated one (the lighter and the darker regions, respectively). The (a) - (d) images correspond approximately to the sample's conductivity marked by the letters "a" – "d" in Fig. 1. Samples completely coated with SAM, as well as untreated crystals, appear uniform in SEM.

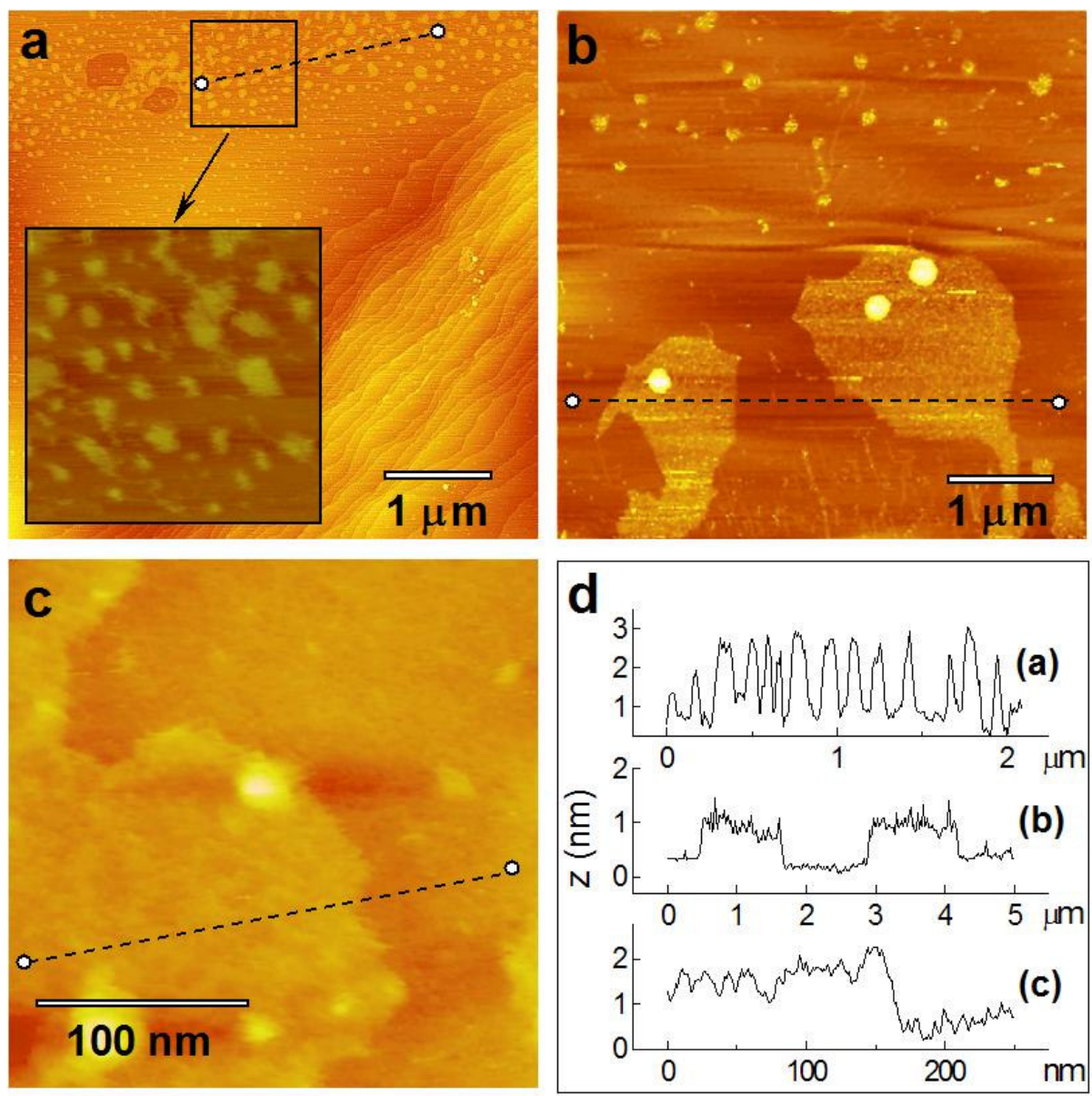

**Fig. 3.** Atomic force microscopy images of an (*a*,*b*)-facet of rubrene single crystal partially coated with (tridecafluoro-1,1,2,2-tetrahydrooctyl)trichlorosilane SAM. (**a**) 5μm x 5μm scan, showing both the SAM nucleation regions (the "patches" at the top), and rubrene molecular steps (the thin wavy lines in the lower right corner). The inset is a higher resolution scan (1μm x 1μm) of the nucleation regions. (**b**) 5μm x 5μm scan, showing larger SAM regions at a molecularly flat rubrene terrace. (**c**) High resolution 250nm x 250nm scan, showing the boundary of a SAM island. (**d**) The AFM profiles taken across the dotted lines in **a**, **b**, and **c**, indicating that the average height of the SAM layer is 1.3 ± 0.2 nm, consistent with the length of the SAM molecule.

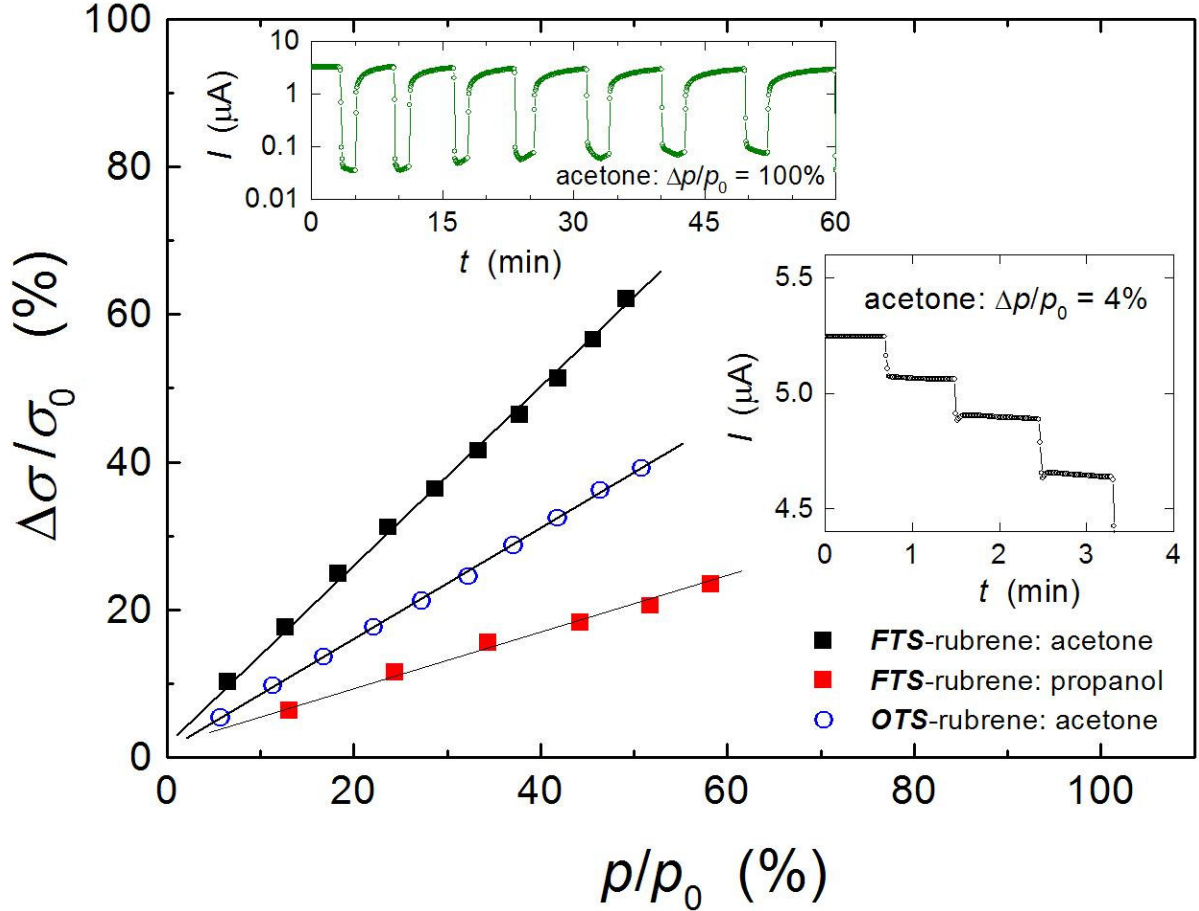

**Fig. 4.** Relative change of conductivity of SAM-functionalized rubrene, $\Delta\sigma/\sigma_0$, as a function of the relative vapor density of acetone and propanol, $p/p_0$, shown for the case of two SAMs on rubrene surface: (tridecafluoro-1,1,2,2-tetrahydrooctyl)trichlorosilane (FTS) and n-octadecyltrichlorosilane (OTS). The top inset shows a semilog $I(t)$ plot for FTS-rubrene sample, periodically exposed to saturated fumes of acetone: a large, very fast and completely reversible effect is observed. The lower inset shows step-wise changes of the current through FTS-functionalized rubrene, when it is exposed to diluted acetone fumes with discretely increasing vapor density (each step corresponds to $\Delta p/p_0 = 4\%$ increase of the vapor density in the test chamber). The sample bias voltage was 5 V. It is evident that the sensor response is a linear function of the analyte density, with the slope dependent on the SAM and the analyte.

**SUPPORTING ONLINE MATERIALS (SOM):**

## A1. Analysis of the effect of different silane and non-silane molecules on conductivity of organic molecular crystals:

Understanding the mechanism of SAM formation at the surface of organic semiconductors requires further studies. The reactions involved in the self-assembly of *trichlorosilanes* at the surface of *silicone oxide* ($SiO_2$) are reasonably well understood. The latter process is based on *surface hydrolysis* of the silane $SiCl_3$ groups that are converted to silanols $Si(OH)_3$ in the presence of water (Sagiv, J., Organized monolayers by adsorption: formation and structure of oleophobic mixed monolayers on solid surfaces. *J. Am. Chem. Soc*. **102**, 92 (1980)). This reaction can occur even in an anhydrous environment and requires only a monolayer of water at the substrate. Further, interactions of Si-OH groups of the hydrolyzed SAM molecule with Si-OH groups of $SiO_2$ substrate result in covalent Si-O-Si bonds. It is believed that not all of the three OH groups of the molecule create covalent Si-O-Si bonds with the substrate. Some of them might form Si-O-Si bonds with the adjacent SAM molecules, forming a planar polymerization along the surface. An ideal model consists of the SAM molecules bonded covalently both to the substrate and to the adjacent SAM molecules. A realistic structure, however, is believed to contain both covalent and hydrogen bonds to the substrate and to the adjacent organosilane molecules. The availability of three OH groups on Si atom increases the probability of forming at least one covalent bond to $SiO_2$ substrate for each molecule.

In our experiment, a persistent and stable effect on the conductivity of molecular crystals has been observed with trichlorosilane molecules. The largest increase of conductivity has been obtained with fluorinated FTS molecules (#1 in the Table A1). It is not evident *a priori* that the $SiCl_3$ group is responsible for binding these SAMs to the surface of organic semiconductors. An alternative bonding mechanisms could be based, for instance, on *dipolar interactions* between the molecules of the organic semiconductor and SAM molecules that have a large built-in dipole moment. Such a mechanism would imply, however, that SAM molecules are attached to the organic surface with their fluoroalkyl tails, because of the *p*-type conductivity of the studied materials. (Our preliminary measurements of vacuum-gap OFETs, in which the channel is treated with FTS SAMs, confirm that SAM-induced surface conductivity is of *p*-type). In order to rule out such mechanism of surface attachment, we have investigated the effect of two types of non-silane molecules on σ: (1) the molecules that have a fluoroalkyl tail and a non-fluorinated head group different from $SiCl_3$ or $Si(OH)_3$, which, similarly to FTS, have a large permanent dipole moment, and (2) fully fluorinated and symmetric molecules that have zero dipole moment. The latter molecules had to be tested to investigate the possibility of surface bonding due to *induced dipole interaction* that does not require a permanent dipole moment.

**Table A1.**

| | Chem. Structure | Chem. Formula/Name | Acronym | Effect on σ |
|---|---|---|---|---|
| 1 | 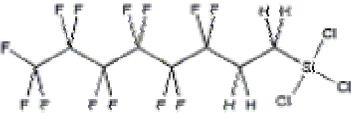 | $C_8H_4Cl_3F_{13}Si$ (tridecafluoro-1,1,2,2-tetrahydrooctyl)trichlorosilane | FTS | very large increase (persistent) |
| 2 | 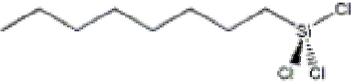 | $C_8H_{17}Cl_3Si$ n-octyltrichlorosilane | OLTS | medium increase (persistent) |
| 3 | 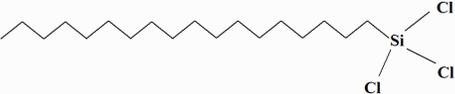 | $C_{18}H_{37}Cl_3Si$ n-octadecyltrichlorosilane | OTS | large increase (persistent) |
| 4 | 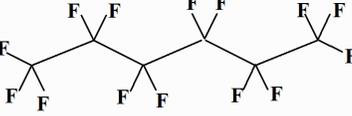 | $CF_3(CF_2)_4CF_3$ tetradecafluorohexane | | none |
| 5 | 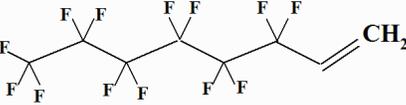 | $CF_3(CF_2)_5CH=CH_2$ 1H,1H,2H-perfluoro-1-octene | | none |
| 6 | 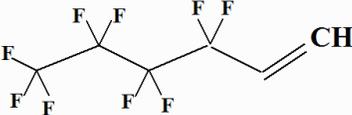 | $CF_3(CF_2)_3CH=CH_2$ 1H,1H,2H-perfluoro-1-hexene | | none |
| 7 | 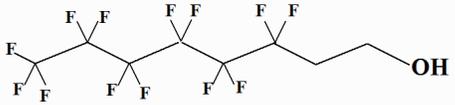 | $CF_3(CF_2)_5CH_2CH_2OH$ 1H,1H,2H,2H-perfluoro-1-octanol | | large increase, but vanishes with vapor removal by pumping |

We have treated our samples with fumes of tetradecafluorohexane (item 4 in the Table below) - a symmetric, highly fluorinated molecule. The test gave negative results: the low initial conductivity of the crystals has not been affected. This indicates that induced dipole interaction is unlikely to be responsible for the binding.

To test asymmetric molecules with a large permanent dipole moment we have chosen the molecules 5, 6 and 7 (Table A1) due to their similarity to FTS molecule. The fumes of 5 and 6 did not have any effect on σ of organic crystals, indicating that a large permanent dipole moment is not a prerequisite for binding to organic surfaces, which also rules out the possibility of SAM attachment to the organics by the fluoroalkyl (or alkyl) tail.

Molecule 7 resembles 5 and 6 in terms of the structure of the fluorinated tail, but has a C-OH head group. This molecule is the only non-trichlorosilane molecule in the list of tested compounds that has resulted in a large increase of the crystal conductivity. The effect, however, fully diminishes with pumping the test chamber by a mechanical pump (i.e., after removing the vapor). This indicates that these molecules physisorb at the surface, suggesting that the C-OH group is able to produce at least a weak hydrogen bond to organic surfaces.

The above analysis strongly suggests that the $SiCl_3$ groups of trichlorosilane molecules are the groups responsible for binding SAM molecules to the surface of organic semiconductors. The microscopic mechanism of this binding and the reactions involved in it are not yet understood. It is possible that the process of surface hydrolysis, similar to that involved in a covalent bonding of trichlorosilane SAMs to the surface of $SiO_2$, plays a key role in this case as well. Possibility of other, non-covalent, types of bonding, such as the hydrogen bonds, have to be investigated. The type and strength of the surface bonding can be studied by thermal desorption measurements, IR spectroscopy and medium energy ion scattering: these experiments are in progress in our laboratory.

**A2. The response of SAM-functionalized organic samples to vapors of different solvents.**

We have performed experiments on the sensitivity of SAM-functionalized organic semiconductors to fumes of different solvents using rubrene and tetracene single-crystal samples functionalized with (tridecafluoro-1,1,2,2-tetrahydrooctyl)trichlorosilane (FTS), n-octyltrichlorosilane (OLTS) and n-octadecyltrichlorosilane (OTS) monolayers (molecules 1-3 in the Table A1).

Prior to the SAM growth, electrical contacts to the organic crystals are prepared either by thermal evaporation of silver through a shadow mask or by painting an aqueous solution of colloidal graphite on the crystal surface. During the silver deposition special care must be taken to minimize the density of defects and contamination of the surface of the crystal. First of all, the high-vacuum gauge must be turned off for the entire duration of pumping the chamber and Ag deposition. This is necessary to prevent the "gauge effect" that results in a formation of traps at organic surfaces as a result of active species generated by high-vacuum gauges or hot filaments in a high vacuum environment[15]. The base pressure $P_0$ in the chamber must be calibrated before the deposition, and it should be low: in our chamber $P_0 = 10^{-7}$ Torr (with a filled liquid $N_2$ trap) after 5-8 hours of pumping. The low base pressure during the deposition will help to minimize the "gauge effect" caused by the hot surface of the resistively heated evaporation source. It will also help to minimize contamination of the channel with evaporated silver atoms that are able to get under the shadow mask after being scattered off the residual molecules in the chamber (Podzorov, V., *et al*., Single-crystal organic field-effect transistors with the hole mobility ~ 8 cm$^2$/Vs. *Appl. Phys. Lett*. **83**, 3504 (2003)).

Figure SOM1 and Fig. 4 (top inset) show that when the samples are exposed to saturated vapors of acetone, propanol, xylene, water and ethanol, a large and reversible changes of the sample's electrical conductivity occur. The effect is rather easy to measure, due to the large conductivity per square of the FTS-functionalized organic crystals.

Exposure to polar analytes results in a decrease of conductivity; in contrast, xylene fumes cause a 75% increase of σ (Fig. SOM1(a)). Sensitivity to $H_2O$ vapor has been observed in the samples with silver contacts, but not in devices with graphite contacts (the case of FTS-functionalized tetracene crystal with Ag contacts is shown in Fig. SOM1(b)). This indicates that dipolar interactions of analyte molecules with the SAM, which are likely to be responsible for the observed sensing effects, could result not only in modification of the channel properties, but also influence the Schottky barriers of the contacts. It remains to be studied in detail why $H_2O$ produces a large effect on SAM-functionalized organic samples with silver contacts, but not on those with the graphite contacts.

Fume delivery to the sample and fume removal have been done by two methods: (1) using a vacuum chamber with a pump and a compartment with a liquid analyte, separated from the chamber by a valve, or (2) by a continuous flow of a mixture of an analyte vapor and nitrogen gas (or pure $N_2$ gas) over the samples. Figure SOM1(c) shows the results of such an experiment, in which a mixture of dry $N_2$ and ethanol is flown through the test chamber with an FTS-functionalized rubrene; the recovery is achieved by purging the chamber with pure $N_2$ gas. In this case, the sensing effect also occurs very quickly; the recovery is somewhat slower than in the case of pumping the fumes out with a mechanical pump, as expected. Nevertheless, a complete recovery is achieved within a matter of minutes.

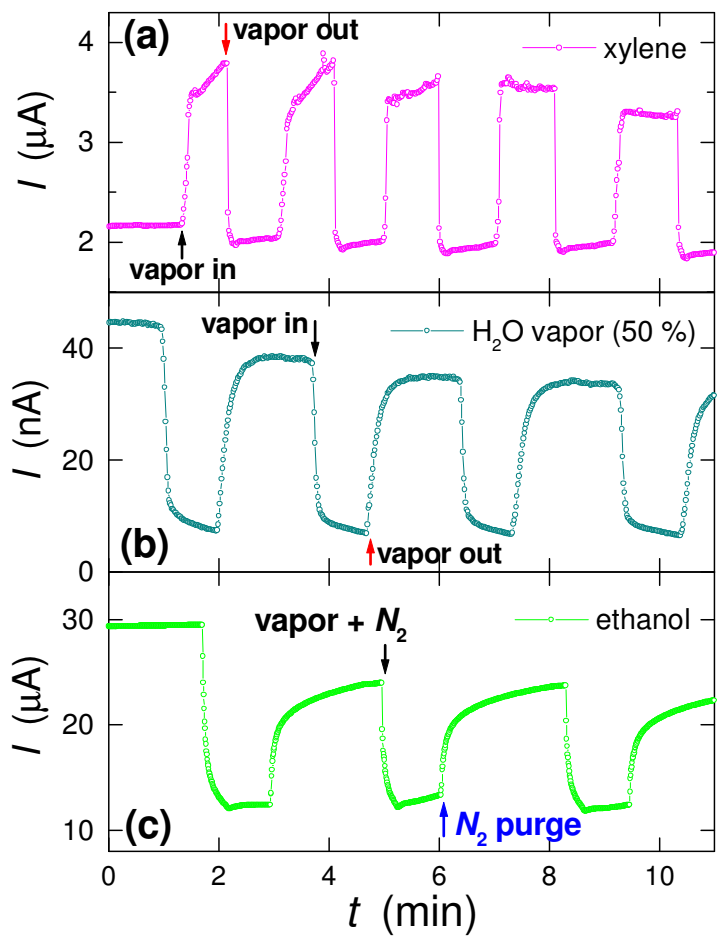

**Fig. SOM1.** Sensing of different analytes with SAM-functionalized rubrene and tetracene crystals. $I(t)$ curves recorded during the multiple exposures of the samples to (**a**) saturated vapor of xylene, (**b**) water vapor (density: 50% of the saturated $H_2O$ vapor at room temperature), (**c**) saturated vapor of ethanol. $I(t)$ curve for acetone is shown in Fig. 4 in the main text. The samples in (a) and (c) are FTS-functionalized rubrene crystals with graphite contacts; the sample in (b) is FTS-functionalized tetracene crystal with thin-film silver contacts. The arrows indicate the time of vapor introduction into the test chamber and vapor removal: in (a) the fumes are removed by pumping the chamber, in (b) and (c) the vapor is introduced by flowing a mixture of the vapor with a dry nitrogen gas, and it is removed by flowing pure nitrogen gas over the samples at a rate of 6 cc/s.

# REFERENCES:

1. Love, J. C., Estroff, L. A., Kriebel, J. K., Nuzzo, R. G. & Whitesides, G. M. Self-assembled monolayers of thiolates on metals as a form of nanotechnology. *Chem. Rev.* **105**, 1103 (2005).
2. Campbell, I. H., Kress, J. D., Martin, R. L., Smith, D. L., Barashkov, N. N. & Ferraris, J. P. Controlling charge injection in organic electronic devices using self-assembled monolayers. *Appl. Phys. Lett.* **71**, 3528 (1997).
3. Akkerman, H. B., Blom, P. W. M., de Leeuw, D. M. & de Boer, B. Towards molecular electronics with large-area molecular junctions. *Nature* **441**, 69 (2006).
4. Heimel, G., Romaner, L., Brédas, J.-L. & Zojer, E. Interface energetics and level alignment at covalent metal-molecule junctions: π-conjugated thiols on gold. *Phys. Rev. Lett.* **96**, 196806 (2006).
5. Briseno, A. L. *et al*. Patterning organic single-crystal transistor arrays. *Nature* **444**, 913 (2006).
6. Kobayashi, S. et al. Control of carrier density by self-assembled monolayers in organic field-effect transistors. *Nature Mater.* **3**, 317 (2004).
7. Facchetti, A., Yoon, M.-H. & Marks, T. J. Gate dielectrics for organic field-effect transistors: new opportunities for organic electronics. *Adv. Mater.* 17, 1705 (2005).
8. Chua, L. L., Zaumseil, J., Chang, J. F., Ou, E. C. W., Ho, P. K. H., Sirringhaus, H. & Friend, R. H. General observation of n-type field-effect behaviour in organic semiconductors. *Nature* **434**, 194 (2005).
9. de Boer, R. W. I., Iosad, N. N., Stassen, A. F., Klapwijk, T. M., and Morpurgo, A. F. Influence of the gate leakage current on the stability of organic single-crystal field-effect transistors. *Appl. Phys. Lett.* **86**, 032103 (2005).
10. Hulea, I. N., Fratini, S., Xie, H., Mulder, C. L., Iosad, N. N., Rastelli, G., Ciuchi, S. & Morpurgo, A. F. Tunable Fröhlich polarons in organic single-crystal transistors. *Nature Mater.* **5**, 982 (2006).
11. Panzer, M. J. & Frisbie, C. D. High charge carrier densities and conductance maxima in single-crystal organic field-effect transistors with a polymer electrolyte gate dielectric. *Appl. Phys. Lett.* **88**, 203504 (2006).
12. de Boer, R. W. I., Gershenson, M. E., Morpurgo, A. F., & Podzorov, V. Organic single-crystal field-effect transistors. *Phys. Stat. Solidi*, **201**, 1302 (2004).
13. Gershenson, M. E., Podzorov, V. & Morpurgo, A. F. *Colloquium*: Electronic transport in single-crystal organic transistors. *Rev. Mod. Phys.* **78**, 973 (2006).
14. The organosilane compounds have been purchased from Gelest.
15. Podzorov, V., *et al*., Interaction of organic surfaces with active species in the high-vacuum environment. *Appl. Phys. Lett.* **87**, 093505 (2005).
16. Podzorov, V. *et al*., In preparation.
17. Sugimura, H., Hayashi, K., Saito, N., Nakagiri, N. & Takai, O. Surface potential microscopy for organized molecular systems. *Appl. Surf. Sci.* **188**, 403 (2002).
18. Crone, B., Dodabalapur, A., Gelperin, A., Torsi, L., Katz, H. E., Lovinger, A. & Bao, Z. Electronic sensing of vapors with organic transistors. *Appl. Phys. Lett.* **78**, 2229 (2001).